\documentclass[prb,twocolumn,showpacs]{revtex4}
\usepackage{graphicx} 
\newcommand{\etal}{{ \it et al. }}

\usepackage{amsmath,amsthm,amssymb,mathrsfs}
\usepackage{bm}

\begin{document}

\title{Thermal switching rate of a ferromagnetic material with uniaxial anisotropy}

\author{Tomohiro Taniguchi and Hiroshi Imamura}
 \affiliation{
 Spintronics Research Center, National Institute of Advanced Industrial Science and Technology, 
 Tsukuba, Ibaraki 305-8568, Japan
 }

 \begin{abstract}
  { 
    The field dependence of the thermal switching rate of a ferromagnetic material 
    with uniaxial anisotropy was studied by solving the Fokker-Planck equation. 
    We derived the analytical expression of the thermal switching rate 
    using the mean first-passage time approach, 
    and found that Brown's formula [Phys. Rev. {\bf 130}, 1677 (1963)] is applicable even in the low barrier limit 
    by replacing the attempt frequency with the proper factor which is expressed by the error function. 
  }
 \end{abstract}

 \maketitle


\section{Introduction}
\label{sec:Introduction}

The thermal switching rate is an important quantity of ferromagnetic materials 
for applications in spintronics devices such as 
a magnetic recording media and spin random access memory (Spin RAM). 
The smaller the thermal switching rate, 
the longer the data retention period.
In most experiments \cite{hayakawa08,yakata09,yakata10}, 
the thermal switching rate is obtained by 
measuring the magnetization switching probability 
in the thermally activated region 
with the assistance of the applied field $H_{\rm appl}$ 
or electric current $I$ (spin torque \cite{slonczewski96,berger96}).
Then the thermal switching rate can be analyzed by Brown's formula
\cite{brown63} which is given by 
\begin{equation}
  R_{1} 
  = 
  \frac{\alpha\gamma H_{\rm K}}{1+\alpha^{2}}
  \sqrt{
    \frac{\Delta_{0}}{\pi}
  }
  \left(
    1 - h^{2}
  \right)
  \left(
    1 - h
  \right)
  e^{-\Delta_{0}(1-h)^{2}}, 
  \label{eq:switching_rate_Brown}
\end{equation}
where $\alpha$, $\gamma$, and $H_{\rm K}$ are 
the Gilbert damping constant, gyromagnetic ratio, and uniaxial anisotropy field, respectively. 
$\Delta_{0}=MH_{\rm K}V/(2k_{\rm B}T)$ is the thermal stability, 
where $M$, $V$, and $T$ are the magnetization, volume of the free layer, and temperature, respectively. 
$h=-H_{\rm appl}/H_{\rm K}$ is the ratio of the applied field to the uniaxial anisotropy field, and 
$\Delta_{0}(1-h)^{2}$ is the barrier height of the magnetic free energy. 


Brown's formula was derived in the high barrier limit
$\Delta_{0}(1-h)^{2} \gg 1$.  
In this case, it takes a long time to observe the thermal switching 
(more than a day or a year, depending on the value of $\Delta_{0}$).  
On the other hand, the most interesting and important limit in the experiments is 
the low barrier limit with high thermal stability \cite{comment1} 
because the experimental determination of the switching rate, 
as well as the thermal stability, requires 
a large number of observations of switching: 
For example, in Ref. \cite{yakata09}, 
4000 times of switching were observed for one sample.  
Thus, in the experiments, a high field $|h| \lesssim 1$ is applied 
to the ferromagnetic materials 
to quickly observe the thermal switching.  
In such limits, Brown's formula gives unphysical predictions, as shown in this paper, 
and is no longer applicable.
However, Brown's formula has been widely used in experiments to determine 
the switching rate and thermal stability \cite{hayakawa08,yakata09,yakata10}, 
which may, for example, lead to a significant error in the evaluation of the retention period of Spin RAM. 
Thus, it is important to study the validity of Brown's formula 
in the low barrier limit with high thermal stability. 
Also, the derivation of a simple and useful analytical expression 
of the thermal switching rate is desirable 
for the experiments. 


In this paper, we studied the field dependence of the thermal switching rate of 
ferromagnetic materials with uniaxial anisotropy 
by solving the Fokker-Planck equation. 
We investigated the analytical expression of the mean first-passage time, 
and found that Brown's formula is applicable 
even in the low barrier limit 
by replacing the attempt frequency with the proper factor, 
which is expressed by the error function.  
We also compared the analytical formulas in several limits 
with the numerically calculated values
and confirmed the validity of each analytical formula.


The paper is organized as follows. 
In Sec. \ref{sec:Mean First-Passage Time of Magnetization Switching}, 
we derive the theoretical expression of the switching rate of the uniaxially anisotropic ferromagnetic material 
by the mean first-passage time approach, 
and show that the mean first-passage time approach is reduced to Brown's formula 
in the high barrier limit. 
Sections \ref{sec:Analytical Expressions of Switching Rate in Other Limits} and 
\ref{sec:Comparison with Numerical Calculation} are the main parts of this paper. 
We derive the analytical formulas of the switching rate 
in the low barrier and low thermal stability limits 
and compare these to the numerically calculated switching rate. 
Section \ref{sec:Conclusions} is devoted to the conclusions. 


\section{Mean First-Passage Time of Magnetization Switching}
\label{sec:Mean First-Passage Time of Magnetization Switching}

We first derive an analytical expression of the mean first-passage time of 
the ferromagnetic material with uniaxial anisotropy by solving the Fokker-Planck equation. 
We assume that the dynamics of the magnetization in the ferromagnetic material is aptly described 
by the Landau-Lifshitz-Gilbert equation given by
\begin{equation}
  \frac{d\mathbf{m}}{d t}
  =
  -\gamma
  \mathbf{m}
  \times
  \mathbf{H}
  +
  \alpha
  \mathbf{m}
  \times
  \frac{d\mathbf{m}}{d t},
  \label{eq:LLG}
\end{equation}
where $\mathbf{m}=(\sin\theta\cos\varphi,\sin\theta\sin\varphi,\cos\theta)$ is the unit vector 
pointing to the direction of the magnetization $\mathbf{M}$. 
The magnetic field $\mathbf{H}=-\partial F/\partial(\mathbf{M}V)=(H_{\rm appl}+H_{\rm K}\cos\theta)\mathbf{e}_{z}$ consists of 
the applied field and the uniaxial anisotropy field.  
$F=-MH_{\rm appl}V\cos\theta-(MH_{\rm K}V/2)\cos^{2}\theta$ is the magnetic free energy.
In the uniaxially anisotropic system, the magnetization dynamics is described by $\theta$ alone.  
We assume that $|h|<1$ because we are interested in the thermally activated region. 
Then, the free energy has two local minima at $\theta=0,\pi$ 
and a maximum at $\theta_{\rm m}=\cos^{-1}h$.
The difference of the free energies at $\theta=0$ and $\theta=\theta_{\rm m}$ 
(i.e., the barrier height) is given by $\Delta_{0}(1-h)^{2}$. 
Since below we focus on the thermal switching from $\theta=0$ to $\theta=\pi$, 
we assume that the field is applied to the $-z$ direction (i.e., $h>0$).


At finite temperature, the thermal fluctuation gives 
additional torque on the magnetization, which is described by $-\gamma
\mathbf{m} \times \mathbf{h}$.  Here the random field $\mathbf{h}$
satisfies the fluctuation-dissipation theorem,
\begin{equation}
  \langle h_{i}(t) h_{j}(t^{\prime}) \rangle 
  =
  \frac{2\alpha k_{\rm B}T}{\gamma MV}
  \delta_{ij}
  \delta(t-t^{\prime}),
  \label{eq:FDT}
\end{equation}
where $i,j=x,y,z$ and $T$ is the temperature. 
$\langle \cdots \rangle$ means the statistical average. 
The stochastic motion of the magnetization due to the thermal
fluctuation is described by the probability function
$P(\theta,t|\theta^{\prime},t^{\prime})$, which represents the
transition probability from the state $\theta^{\prime}$ at time $t^{\prime}$ to
the state $\theta$ at time $t$.  From Eq. (\ref{eq:LLG}), the
Fokker-Planck equation for the probability function is given by \cite{brown63}
\begin{equation}
\begin{split}
  \frac{\partial P}{\partial t}
  =
  \alpha
  \gamma^{\prime}
  \frac{\partial}{\partial\theta}
  &
  \left[
    \left(
      H_{\rm appl}
      +
      H_{\rm K}
      \cos\theta
    \right)
    \sin\theta
    -
    \frac{k_{\rm B}T}{MV}
    \cot\theta
  \right] P
\\
  &+
  \frac{\alpha\gamma^{\prime} k_{\rm B}T}{MV}
  \frac{\partial^{2}P}{\partial\theta^{2}},
  \label{eq:forward_FP}
\end{split}
\end{equation}
or in terms of $W=P/\sin\theta$, 
\begin{equation}
\begin{split}
  \frac{\partial W}{\partial t}
  \!=\!
  \frac{\alpha\gamma^{\prime}}{\sin\theta}
  \frac{\partial}{\partial\theta}
  \!\!
  &
  \left\{
  \!
    \sin\theta
    \!\!
    \left[
    \!
      \left(
        H_{\rm appl}
        \!+\!
        H_{\rm K}
        \!
        \cos\theta
      \right)
      \!
      \sin\theta
      W
      \!+\!
      \frac{k_{\rm B}T}{MV}
      \frac{\partial W}{\partial\theta}
    \right]\!
  \right\},
  \label{eq:FP}
\end{split}
\end{equation}
where $\gamma^{\prime}=\gamma/(1+\alpha^{2})$.  
Brown \cite{brown63} calculated the switching rate 
by approximately solving Eq. (\ref{eq:FP}) 
or by investigating the smallest nonvanishing eigenvalues,
which characterize the time relaxation of the probability function. 
In this paper, 
we employ the backward Fokker-Planck approach to derive the mean first-passage time approach, 
which is the same as the eigenvalues of Eq. (\ref{eq:FP}) 
and is useful to evaluate the thermal switching rate in several limits. 


The mean first-passage time $\mathcal{T}(\theta^{\prime})$ \cite{hanggi90,gardiner83}, 
which characterizes how long the magnetization in the potential $F$ stays within
the region $0 \le \theta \le \theta_{\rm m}$, is defined as
\begin{equation}
  \mathcal{T}(\theta^{\prime})
  =
  \int_{0}^{\infty} d t
  \int_{0}^{\theta_{\rm m}} d\theta
  P(\theta,t|\theta^{\prime},0).
\end{equation}
The equation to determine the mean first-passage time is obtained from 
the backward Fokker-Planck equation \cite{hanggi82} as
\begin{equation}
\begin{split}
  \frac{\partial P}{\partial t^{\prime}}
  =&
  \alpha
  \gamma^{\prime}
  \left[
    \left(
      H_{\rm appl}
      \!+\!
      H_{\rm K}
      \cos\theta^{\prime}
    \right)
    \sin\theta^{\prime}
    \!-\!
    \frac{k_{\rm B}T}{MV}
    \cot\theta^{\prime}
  \right]
  \frac{\partial P}{\partial\theta^{\prime}}
\\
  &-
  \frac{\alpha\gamma^{\prime}k_{\rm B}T}{MV}
  \frac{\partial^{2}P}{\partial\theta^{\prime 2}},
\end{split}
\end{equation}
and is given by 
\begin{equation}
\begin{split}
  \frac{d^{2}\mathcal{T}}{d\theta^{\prime 2}}
  &+
  \left[
    \cot\theta^{\prime}
    \!-\!
    \frac{MV}{k_{\rm B}T}
    \left(
      H_{\rm appl}
      \!+\!
      H_{\rm K}
      \cos\theta^{\prime}
    \right)
    \sin\theta^{\prime}
  \right]
  \frac{d\mathcal{T}}{d\theta^{\prime}}
\\
  &=
  -\frac{2\Delta_{0}}{\alpha\gamma^{\prime} H_{\rm K}}.
  \label{eq:FP_T1}
\end{split}
\end{equation}
To solve Eq. (\ref{eq:FP_T1}), 
it is convenient to transfer the variable $\theta^{\prime}$ to $z=\cos\theta^{\prime}$. 
Then, Eq. (\ref{eq:FP_T1}) can be expressed as
\begin{equation}
\begin{split}
  (1\!-\!z^{2})
  \frac{d^{2}\mathcal{T}}{d z^{2}}
  &\!-\!
  2
  \left[
    z
    \!+\!
    \Delta_{0}
    (1-z^{2})
    \left(
      h
      \!-\!
      z
    \right)
  \right]
  \frac{d\mathcal{T}}{d z}
  =
  -\frac{2\Delta_{0}}{\alpha\gamma^{\prime} H_{\rm K}}.
  \label{eq:FP_T2}
\end{split}
\end{equation} 
The mean first-passage time is obtained by solving Eq. (\ref{eq:FP_T1}) 
with the appropriate boundary conditions. 
We use the reflecting and absorbing boundary conditions 
at $\theta^{\prime}=0$ and $\theta^{\prime}=\theta_{\rm m}$, respectively \cite{gardiner83}, 
that is, $(d\mathcal{T}/d\theta^{\prime})_{\theta^{\prime}=0}=0$ and $\mathcal{T}(\theta_{\rm m})=0$. 
Then the mean first-passage time is given by 
\begin{equation}
\begin{split}
  \mathcal{T}(z)
  =
  \frac{2\Delta_{0}}{\alpha\gamma^{\prime} H_{\rm K}} 
  &
  \int_{h}^{z} d x 
  \int_{x}^{1} d y 
  \frac{\exp\{-\Delta_{0}[(x-h)^{2}-(y-h)^{2}]\}}{1-x^{2}}.
  \label{eq:MFPT}
\end{split}
\end{equation}
Here, $x$ and $y$ are the integration variables. 
Equation (\ref{eq:MFPT}) was derived in Ref. \cite{coffey98}. 
Once the magnetization reaches $\theta=\theta_{\rm m}$, 
it moves to $\theta<\theta_{\rm m}$ or $\theta>\theta_{\rm m}$ with the probability $1/2$. 
Thus, the switching rate is given by \cite{hanggi90,coffey98}
\begin{equation}
  R
  =
  \frac{1}{2\mathcal{T}(z=1)}.
\end{equation}
Below, we denote $\mathcal{T}(z=1)$ as $\mathcal{T}$. 
For the later discussion, 
we also define the normalized (dimensionless) switching rate as
\begin{equation}
  r
  =
  \frac{2\Delta_{0}}{\alpha\gamma^{\prime}H_{\rm K}} R.
  \label{eq:normalized_switching_rate}
\end{equation}


It should be noted that the $y$ integral in Eq. (\ref{eq:MFPT}) can be written in a different form 
by introducing the imaginary error function ${\rm erfi}(z)=(2/\sqrt{\pi})\int_{0}^{z} d y e^{y^{2}}$, 
that is, 
\begin{equation}
\begin{split}
  &
  \int_{x}^{1} d y 
  \exp
  \left[
    \Delta_{0}
    (y-h)^{2}
  \right]
\\
  &=
  \frac{1}{2}
  \sqrt{
    \frac{\pi}{\Delta_{0}}
  }
  \left[
    {\rm erfi}
    \left(
      \sqrt{\Delta_{0}}
      (1-h)
    \right)
    -
    {\rm erfi}
    \left(
      \sqrt{\Delta_{0}}
      (x-h)
    \right)
  \right].
  \label{eq:IEF}
\end{split}
\end{equation}
By using the Taylor expansion of the imaginary error function around $x=1$, 
we found that the mean first-passage time can be expressed as 
\begin{equation}
  \mathcal{T}
  =
  \frac{2\Delta_{0}}{\alpha\gamma^{\prime} H_{\rm K}}
  \sum_{\ell=1}^{\infty}
  C_{\ell}
  \int_{h}^{1} d x 
  \frac{(1-x)^{\ell-1}}{1+x}
  e^{-\Delta_{0}(x-h)^{2}},
  \label{eq:MFPT_1}
\end{equation}
where the coefficient $C_{\ell}=C_{\ell}(\Delta_{0},h)$ is given by 
\begin{equation}
  C_{\ell}(\Delta_{0},h)
  =
  \frac{(-1)^{\ell-1}}{\ell !}
  \frac{d^{\ell-1}}{d y^{\ell-1}}
  e^{\Delta_{0}(y-h)^{2}}
  \bigg|_{y=1}.
  \label{eq:coff_Cn}
\end{equation}
The first three terms of $C_{\ell}(\Delta_{0},h)$ are given by 
\begin{align}
  &
  C_{1}
  =
  e^{\Delta_{0}(1-h)^{2}},
  \\
  &
  C_{2}
  =
  -\Delta_{0}
  (1-h)
  e^{\Delta_{0}(1-h)^{2}},
  \\
  &
  C_{3}
  =
  \frac{\Delta_{0}}{3}
  \left[
    1
    +
    2 \Delta_{0}
    (1-h)^{2}
    \right]
  e^{\Delta_{0}(1-h)^{2}}.
\end{align}
References \cite{brown63,aharoni69,garanin90,scully92,coffey95,garanin96,klik98,coffey98,coffey01,kalmykov03} also calculate the switching rate 
by analytically or numerically solving the forward or backward Fokker-Planck equation. 
For example, in Refs. \cite{brown63,aharoni69,coffey95,coffey01,kalmykov03}, 
the switching rate is calculated by expanding the probability function with the Legendre polynomials. 
In Ref. \cite{coffey98}, the mean first-passage time 
after the integration shown in Eq. (\ref{eq:IEF}) is derived 
in terms of the Kummer's function,
$M(a,b,z)=\sum_{n=0}^{\infty}[\Gamma(a+n)\Gamma(b) z^{n}]/[n! \Gamma(a)\Gamma(b+n)]$, 
where $\Gamma(z)=\int_{0}^{\infty} dt t^{z-1}e^{-t}$ is the $\Gamma$ function 
[the Kummer's function satisfies the relation 
${\rm erfi}(u)=(2u/\sqrt{\pi})M(1/2,3/2,u^{2})$].
To obtain the analytical formula of the switching rate, 
these studies mainly investigated 
the high barrier ($\Delta_{0}(1-h)^{2} \gg 1$), 
small applied field ($h \simeq 0$), or low thermal stability ($\Delta_{0} \ll 1$) limits. 
Although these constitute exact solutions, 
the particular limit of high fields and high thermal stability treated in
the present manuscript was not singled out for particular attention, 
apart from a few numerical evaluations of 
the mean first-passage time in this regime \cite{coffey98}. 
On the other hand, we note that the expansion shown in Eq. (\ref{eq:MFPT_1}) is useful 
in deriving the analytical formula of the switching rate 
in the limits of $\Delta_{0}(1-h)^{2} \ll 1$, $h \lesssim 1$, and $\Delta_{0} \gg 1$, 
which can be directly applied to analyze the experiments to evaluate the thermal switching rate and thermal stability accurately.

Before proceeding with further calculations, 
let us give a brief comment on the spin transfer torque. 
In a ferromagnetic multilayer with a pinned layer, 
the electric current applied to the multilayer exerts a spin transfer torque \cite{slonczewski96,berger96},
which gives the additional term $\gamma H_{\rm s} \mathbf{m} \times (\mathbf{p} \times \mathbf{m})$ to Eq. (\ref{eq:LLG}). 
Here $H_{\rm s}=\hbar \eta I/(2eMV)$ and $\mathbf{p}$ are 
the strength of the spin transfer torque and the unit vector pointing in 
the direction of the magnetization of the pinned layer, respectively. 
The positive current ($I>0$) with the spin polarization $\eta$ is defined as 
the electron flow from the pinned to the free layer. 
In general, the effect of the spin torque cannot be included into the magnetic free energy, 
and the steady-state solution of the Fokker-Planck equation deviates from the Boltzmann distribution. 
Thus, the theoretical approach to the thermal switching rate shown in Ref. \cite{brown63} is no longer applicable. 
However, in the uniaxially anisotropic system with $\mathbf{p} \parallel \mathbf{e}_{z}$, 
the effect of the spin transfer torque can be taken into account by replacing $H_{\rm appl}$ in the magnetic free energy 
by $H_{\rm appl}+H_{\rm s}/\alpha$ \cite{suzuki09,taniguchi11a,taniguchi11b}. 
Thus, the following discussions are applicable to the spin transfer torque system 
in this special case. 
It should be noted that Suzuki \etal \cite{suzuki09} showed that 
the results in the special case can be applied to general systems 
by replacing the parameters (for example, $H_{\rm K}$) with the appropriate values.

At the end of this section, 
let us show that Eq. (\ref{eq:MFPT}) reproduces Brown's formula 
in the high barrier limit, $\Delta_{0}(1-h)^{2} \gg 1$. 
In this limit,
the integral in Eq. \eqref{eq:MFPT} is dominated by the contribution around 
$x=h$.
Since $\lim_{z \to 0}{\rm erfi}(z)=0$, 
the result of the $y$ integral [Eq. (\ref{eq:IEF})] in the limit of $x \to h$ is given by 
$\sqrt{\pi}{\rm erfi}[\sqrt{\Delta_{0}}(1-h)]/(2 \sqrt{\Delta_{0}})$. 
By using the following formula, 
\begin{equation}
\begin{split}
  \frac{2}{\sqrt{\pi}}
  \int d y 
  e^{y^{2}}
  &=
  \frac{e^{y^{2}}}{\sqrt{\pi}}
  \left(
    \frac{1}{y}
    +
    \frac{1}{2y^{3}}
    +
    \frac{3}{4y^{5}}
    +
    \cdots
  \right)
\\
  &=
  \frac{e^{y^{2}}}{\sqrt{\pi}}
  \sum_{n=0}^{\infty}
  \frac{(2n-1)(2n-3),... 3 \times 1}{2^{n}y^{2n+1}},
\end{split}
\end{equation}
we found that 
\begin{equation}
  \frac{1}{2}
  \sqrt{\frac{\pi}{\Delta_{0}}}
  {\rm erfi}
  \left(
    \sqrt{\Delta_{0}}
    (1-h)
  \right)
  \simeq
  \frac{e^{\Delta_{0}(1-h)^{2}}}{2 \Delta_{0}(1-h)}.
\end{equation}
On the other hand, the $x$ integral is given by 
\begin{equation}
\begin{split}
  \int_{h}^{1} d x 
  \frac{e^{-\Delta_{0}(x-h)^{2}}}{1-x^{2}}
  &\simeq
  \frac{1}{1-h^{2}}
  \int_{0}^{1-h} d u 
  e^{-\Delta_{0}u^{2}}
\\
  &=
  \frac{1}{2}
  \sqrt{
    \frac{\pi}{\Delta_{0}}
  }
  \frac{{\rm erf}[\sqrt{\Delta_{0}}(1-h)]}{1-h^{2}},
\end{split}
\end{equation}
where ${\rm erf}(z)=(2/\sqrt{\pi})\int_{0}^{z} d y e^{-y^{2}}$ is the error function. 
Since $\lim_{z \to \infty}{\rm erf}(z)=1$ 
in the high barrier limit, 
the $x$ integral can be approximated to $\sqrt{\pi}/[2\sqrt{\Delta_{0}}(1-h^{2})]$. 
Then, Eq. (\ref{eq:MFPT}) yields
\begin{equation}
  \mathcal{T}_{1} 
  =
  \frac{1}{2 \alpha \gamma^{\prime} H_{\rm K}}
  \sqrt{
    \frac{\pi}{\Delta_{0}}
  }
  \frac{e^{\Delta_{0}(1-h)^{2}}}{(1-h^{2})(1-h)}. 
  \label{eq:MFPT_Brown}
\end{equation}
The switching rate $R_{1}=1/(2\mathcal{T}_{1})$ is identical to Brown's formula, Eq. (\ref{eq:switching_rate_Brown}). 
We can easily find that $R_{1}$ is zero in the high field limit ($h \to 1$). 
Also, the attempt frequency, which is defined by $R_{1}e^{\Delta_{0}(1-h)^{2}}$, increases with decreasing temperature in the limit of $T \to 0$.
However, physically, the switching rate should increase with increasing field strength,
and the attempt frequency should be zero in the zero temperature limit. 
The origin of these contradictions is the high barrier assumption, $\Delta_{0}(1-h)^{2} \gg 1$, in its derivation. 



\section{Analytical Expressions of Switching Rate in Other Limits}
\label{sec:Analytical Expressions of Switching Rate in Other Limits}

In this section, 
we derive the analytical expressions of the switching rate 
in the low barrier [$\Delta_{0}(1-h)^{2} \ll 1$] and low thermal stability ($\Delta_{0} < 1$) limits. 

First, we investigate the analytical expression of the switching rate 
in the high field limit ($h \to 1$) with a high thermal stability ($\Delta_{0} \gg 1$),
that is, the low barrier limit [$\Delta_{0}(1-h)^{2} \ll 1$]. 
In this limit, it is sufficient to take into account the term of $\ell=1$ in Eq. (\ref{eq:MFPT_1}), 
and the mean first-passage time is given by 
\begin{equation}
  \mathcal{T}_{2}
  \simeq
  \frac{2\Delta_{0}}{\alpha\gamma^{\prime} H_{\rm K}}
  e^{\Delta_{0}(1-h)^{2}}
  \int_{h}^{1} d x 
  \frac{e^{-\Delta_{0}(x-h)^{2}}}{1+x}.
\end{equation}
Although $C_{\ell}$ for $\ell \ge 2$ is larger than unity for $\Delta_{0} \gg 1$, 
we can neglect the higher order terms of $C_{\ell}$ in Eq. (\ref{eq:MFPT_1}) 
because the integral interval, $[h,1]$, is very close to $x=1$, 
and thus, the factor $(1-x)^{\ell-1}$ for $\ell \ge 2$ in Eq. (\ref{eq:MFPT_1}) can be approximated to zero. 
Similarly, the factor $1/(1+x)$ in the $x$ integral can be approximated to $1/2$. 
Then, the switching rate $R_{2}=1/(2\mathcal{T}_{2})$ is obtained as
\begin{equation}
  R_{2}
  \simeq
  \frac{\alpha\gamma^{\prime}H_{\rm K}}{\sqrt{\pi\Delta_{0}}}
  \frac{e^{-\Delta_{0}(1-h)^{2}}}{{\rm erf}[\sqrt{\Delta_{0}}(1-h)]}.
  \label{eq:MFPT_high_field_limit}
\end{equation}
This is the analytical expression of the switching rate in the low barrier (high field) limit, 
and is one of the main findings in this paper. 
It should be noted that $R_{2}$ increases with increasing field, 
and the attempt frequency $R_{2}e^{\Delta_{0}(1-h)^{2}}$ vanishes in the limit of $T\to 0$,
which agrees well with our intuition.
Equation (\ref{eq:MFPT_high_field_limit}) can be applied to the experiments 
to evaluate the thermal switching rate $R_{2}$ and the thermal stability $\Delta_{0}$. 


We also derive the analytical expression of the switching rate 
in the low thermal stability limit ($\Delta_{0} \ll 1$). 
In this limit, $C_{\ell}$ in Eq. (\ref{eq:coff_Cn}) can be approximated to 
$C_{\ell} \simeq \delta_{\ell,1}$. 
Then, Eq. (\ref{eq:MFPT_1}) can be approximated to 
\begin{equation}
  \mathcal{T}_{3}
  \simeq
  \frac{2\Delta_{0}}{\alpha\gamma^{\prime}H_{\rm K}}
  \int_{h}^{1} d x 
  \frac{1}{1+x}.
\end{equation}
Thus, the switching rate $R_{3}=1/(2\mathcal{T}_{3})$ is given by 
\begin{equation}
  R_{3}
  \simeq
  \frac{\alpha\gamma^{\prime}H_{\rm K}}{4 \Delta_{0}}
  \frac{1}{\log[2/(1+h)]}.
  \label{eq:MFPT_Klein}
\end{equation}
The switching rate $R_{3}$ increases 
with increasing field magnitude, 
and vanishes in the limit of $T\to 0$.
It should be noted that Eq. (\ref{eq:MFPT_Klein}) is applicable 
to the whole range of the field, $|h| \le 1$. 
We also noted that Eq. (\ref{eq:MFPT_Klein}) is consistent with the result of Klein 
derived using a different approach \cite{klein52}. 
Coffey \etal \cite{coffey01} also derived Eq. (\ref{eq:MFPT_Klein}) 
in which the concept of the mean first-passage time 
for a spherical domain was first introduced. 


By comparing Eqs. (\ref{eq:switching_rate_Brown}) and (\ref{eq:MFPT_high_field_limit}), 
we found that the switching rate is given by the product of 
the attempt frequency and the exponential factor $e^{-\Delta_{0}(1-h)^{2}}$,
and that Brown's formula is applicable even in the low barrier limit by replacing 
the attempt frequency, $\alpha\gamma^{\prime} H_{\rm K}(1-h^{2})(1-h)\sqrt{\Delta_{0}/\pi}$, 
by $\alpha\gamma^{\prime} H_{\rm K}/[\sqrt{\pi\Delta_{0}}{\rm erf}(\sqrt{\Delta_{0}}(1-h))]$. 
However, such exponential dependence is valid only in the high thermal stability limit: 
In the low thermal stability limit, 
the field dependence of the switching rate is described by the logarithm function $\log(1+h)$, 
as shown in Eq. (\ref{eq:MFPT_Klein}). 



\begin{figure}
  \centerline{\includegraphics[width=1.0\columnwidth]{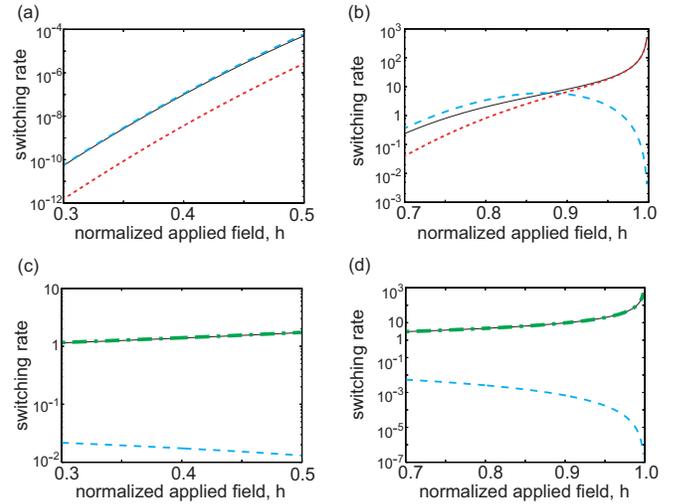}}
  \caption{
           The field $h$ dependences of the normalized switching rates, $\tilde{r}$ and $r_{k}$, 
           where $\tilde{r}$ is the numerical value of Eq. (\ref{eq:normalized_switching_rate})
           while $r_{1}$, $r_{2}$, and $r_{3}$ are 
           Brown's formula [Eq. (\ref{eq:normalized_r1})],
           high field limit [Eq. (\ref{eq:normalized_r2})], 
           and low thermal stability limit [Eq. (\ref{eq:normalized_r3})], respectively. 
           (a), (b) $\tilde{r}$ (solid), $r_{1}$ (dashed), and $r_{2}$ (dotted) in the low and high field region with $\Delta_{0}=60$, respectively. 
           (c), (d) $\tilde{r}$ (solid), $r_{1}$ (dashed), and $r_{3}$ (dashed-dotted) in the low and high field regions with $\Delta_{0}=0.1$, respectively.
  }
  \label{fig:fig1}
\end{figure}



\begin{figure}
  \centerline{\includegraphics[width=1.0\columnwidth]{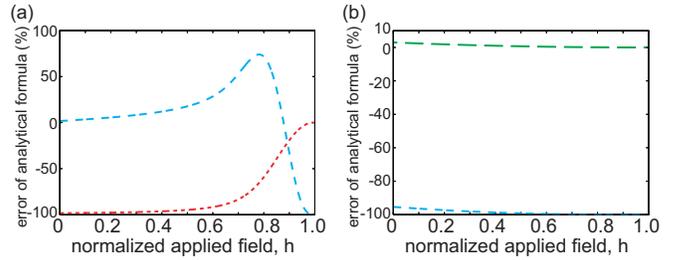}}
  \caption{
           The field $h$ dependences of the errors of the analytical formula, $e_{k}$. 
           (a) $e_{1}$ (dashed) and $e_{2}$ (dotted) with the high thermal stability. 
           (b) $e_{1}$ (dashed) and $e_{3}$ (short dashed-dotted) with the low thermal stability. 
  }
  \label{fig:fig2}
\end{figure}




\section{Comparison with Numerical Calculation}
\label{sec:Comparison with Numerical Calculation}

Finally, we compare the analytical formulas of the switching rate 
[Eqs. (\ref{eq:switching_rate_Brown}), (\ref{eq:MFPT_high_field_limit}), and (\ref{eq:MFPT_Klein})]
with the numerically calculated values of Eq. (\ref{eq:MFPT}). 
For comparison, it is convenient to evaluate the normalized switching rate, Eq. (\ref{eq:normalized_switching_rate}). 
Due to the normalization, 
the switching rates depend on $\Delta_{0}$ and $h$ only. 
We denote the numerical value of Eq. (\ref{eq:normalized_switching_rate}) as $\tilde{r}$, 
and the normalized values of Eqs. (\ref{eq:switching_rate_Brown}), (\ref{eq:MFPT_high_field_limit}), and (\ref{eq:MFPT_Klein}) as 
$r_{1}$, $r_{2}$, and $r_{3}$, respectively. 
Explicitly, these are given by 
\begin{equation}
  r_{1}
  =
  2 
  \sqrt{
    \frac{\Delta_{0}^{3}}{\pi}
  }
  \left(
    1 - h^{2}
  \right)
  \left(
    1 - h
  \right)
  e^{-\Delta_{0}(1-h)^{2}},
  \label{eq:normalized_r1}
\end{equation}
\begin{equation}
  r_{2}
  =
  2 
  \sqrt{
    \frac{\Delta_{0}}{\pi}
  }
  \frac{e^{-\Delta_{0}(1-h)^{2}}}{{\rm erf}[\sqrt{\Delta_{0}}(1-h)]},
  \label{eq:normalized_r2}
\end{equation}
\begin{equation}
  r_{3}
  =
  \frac{1}{2 \log [2/(1+h)]}.
  \label{eq:normalized_r3}
\end{equation}
The normalized values of $r_{2}$ and $r_{3}$ in the limit of $h \to 1$ are identical, 
$1/(1-h)$. 
For the quantitative discussion,
we also define the error of the analytical result by 
$e_{k}=[(r_{k}-\tilde{r})/\tilde{r}]\times 100$ \%. 


Figures \ref{fig:fig1} (a) and (b) are the field dependences of 
$\tilde{r}$ (solid), $r_{1}$ (dashed), and $r_{2}$ (dotted) with the high thermal stability $\Delta_{0}=60$ 
in the low and high field regions, respectively. 
As shown, $r_{1}$ and $r_{2}$ almost overlap with the exact result ($\tilde{r}$) 
in the low and high field regions, respectively. 
For example, the errors of Brown's formula ($r_{1}$) are 
$e_{1}=17.7$ and $-88.4$ \% for $h=0.5$ and $0.95$ 
while those of $r_{2}$ are 
$e_{2}=-94.8$ and $-4.5$ \% for $h=0.5$ and $0.95$, respectively: see Fig. \ref{fig:fig2} (a). 
The good agreement of our formula [Eq. (\ref{eq:MFPT_high_field_limit})] 
in the high field limit is important for the experiments 
to guarantee the validity of Brown's formula. 
Moreover, Brown's formula gives unphysical results in the limit of $h \to 1$;
that is, the switching rate decreases with increasing field, as shown in Fig. \ref{fig:fig1} (b), 
while Eq. (\ref{eq:MFPT_high_field_limit}) gives a physically reasonable 
and quantitatively good estimation of the switching rate. 


In Figs. \ref{fig:fig1} (c) and (d), 
we show the normalized field dependences of $\tilde{r}$ (solid), $r_{1}$ (dashed), and $r_{3}$ (short dashed-dotted) 
with the low thermal stability $\Delta_{0}=0.1$ 
in the low and high field regions, respectively. 
The errors, $e_{1}$ and $e_{3}$, are shown in Fig. \ref{fig:fig2} (b). 
One can clearly see that $r_{3}$ agrees well with the numerical result: 
$e_{3}=0.7$ and $0.007$ \% for $h=0.5$ and $0.95$, respectively. 
On the other hand, Brown's formula reduces 
$e_{1}=-99.2$ and $-99.9$ \% for $h=0.5$ and $0.95$, respectively. 
The large difference between $\tilde{r}$ and $r_{1}$ is due to the break down of the high barrier assumption in Eq. (\ref{eq:switching_rate_Brown}). 


Let us briefly discuss the application of our formula [Eq. (\ref{eq:MFPT_high_field_limit})] to spintronics applications. 
Since the switching rate estimated by Brown's formula is much smaller than the exact value, 
as shown in Fig. \ref{fig:fig1} (b), 
a relatively small value of the thermal stability as a fitting parameter 
is required to fit the experiments with Eq. (\ref{eq:switching_rate_Brown}). 
Then, the estimated value of the retention time of Spin RAM, 
which is proportional to $\exp(\Delta_{0})$, is significantly underestimated. 
On the other hand, Eq. (\ref{eq:MFPT_high_field_limit}) can give accurate estimates of 
the thermal switching rate, the thermal stability, and also the retention time of Spin RAM. 

\section{Conclusions}
\label{sec:Conclusions}

In conclusion, we studied the field dependence of the thermal switching rate 
of a ferromagnet with uniaxial anisotropy theoretically. 
We derived the analytical expression of the switching rate by the mean first-passage time approach, 
and showed that Brown's formula is applicable even in the low barrier limit 
by replacing the attempt frequency with the proper factor 
which is expressed by the error function. 
We also compared the analytical formulas of the switching rate 
for several limits 
with the numerically obtained values 
and showed the validity of each analytical formula. 


\section*{ACKNOWLEDGMENT}
\label{sec:Acknowledgment}

The authors would like to acknowledge 
S. Yuasa, H. Kubota, A. Fukushima, H. Maehara, S. Iba, T. Yorozu, K. Seki, M. Shibata, and Y. Utsumi 
for the valuable discussions they had with us.


\begin{thebibliography}{23}
\bibitem[{com()}]{comment1}
\bibinfo{note}{The low barrier here means a high thermal stability
  $\Delta_{0} \gg 1$ and a high field ($h \simeq 1$), which is different with
  "low-energy-barrier approximation" in Ref. \cite{brown63} where both
  $\Delta_{0}$ and $h$ are small.}
\end{thebibliography}
\end{document}